\documentclass[12pt]{article}
\begin{document}
\title
{Interpolating Action for Strings and Membranes - a Study of Symmetries in
the Constrained Hamiltonian Approach}
\author
{ Rabin Banerjee,\footnote{rabin@bose.res.in}\\
S. N. Bose National Centre for Basic Sciences \\
JD Block, Sector III, Salt Lake City, Kolkata -700 098, India
\and Pradip Mukherjee\\and\\
 Anirban Saha\\
Department of Physics, Presidency College\\
86/1 College Street, Kolkata - 700 073, India}
\maketitle
\abstract
{ A master action for bosonic strings and membranes, interpolating between
the Nambu--Goto and Polyakov formalisms, is discussed. The role of the gauge
symmetries vis-\`{a}-vis reparametrization symmetries of the various actions is
analyzed by a constrained Hamiltonian approach. This analysis reveals the
difference between strings and higher branes, which is essentially tied to a
degree of freedom count. The cosmological term for membranes follows naturally in this scheme. The conncetion of our aproach with the Arnowitt--Deser--Misner representation in general relativity is illuminated.}
\section{Introduction}
String theory was introduced as a candidate for the fundamental theory uniting
all the basic interactions at the Planck scale \cite{pol}. At least five different string theories emerged equally viable, namely the Type I, Type IIA, Type IIB, Heterotic SO(32) and Heterotic $E_8\times E_8$ string theories. This, however, raised doubt about the claimed unique status of string theory. Significant progress in the understanding of different ramifications of string theories has been achieved in the last decade with the discovery of the dualities \cite{du} mapping one theory into another, thereby indicating their essential unity. It is now definitely believed that the different perturbative sectors of string theories occupy different corners of some yet unknown M - theory \cite{mr}. Higher dimensional extended objects like membranes are expected to be instrumental in understanding this new theory. A characteristic feature of these structures is that they are loaded with various symmetries. It is useful to gain an understanding of these symmetries from different points of view.

    Apart from the symmetries of the space - time in which the string ( or the brane ) is embedded, they have diffeomorphism ( diff.) invariance which arises from reparametrization invariance of the world `volume'. The latter can be considered as gauge symmetries implemented by the first class constraints of the theory. The constrained Hamiltonian analysis due to Dirac \cite{dir} is a natural methodology for such problems. It is known that the analysis of space - time symmetries of different field - theoretic models is done by this method in a gauge independent setting \cite{hrt,rb}, a fact demonstrated by numerous applications in the literature \cite{our}. The investigation of the diff. invariance and the gauge symmetries, including their correspondence, from a gauge-independent constrained Hamiltonian approach is therefore strongly suggested. In the present paper we will give a detailed and comprehensive analysis of the symmetries of strings and membranes from this approach. Since the essential features of the diff. invariance are contained in the  bosonic version of the strings (or membranes) we will only consider such models. 

       The action for a string can be chosen in analogy with the relativistic particle as the proper area of the world sheet swept out by the dynamical string. This gives the Nambu--Goto ( NG ) formalism which, however, poses problems in quantization. A redundant description, where the world sheet metric coefficients are considered as independent fields, has been shown by Polyakov to be particularly suitable in this context. The ensuing action is known as the Polyakov action. The equivalence between the two approaches can be established on shell by solving the independent metric in the Polyakov action. The classical correspondence is assumed to lead to equivalent results at the quantum level \cite{pol}. Understanding this correspondence from different viewpoints will, naturally, be useful. Again, the Nambu--Goto and the Polyakov actions have their counterparts for the dynamical membranes \cite{t}. However, there is a distinctive feature of the Polyakov actions for the higher branes as opposed to the strings. This originates from the difference between the number of independent metric elements with the number of independent reparametrizations. The three components of the string world sheet metric are all fixed; two by the reparametrization symmetries and one by the Weyl invariance. For the membrane which has six components of the metric, only three are fixed by reparametrization symmetry while three are indeterminable. We show that it is precisely this freedom which leads to the emergence of the cosmological term for the membrane action.

     A deeper connection between the two forms for the string action has been
demonstrated in \cite{rbs} by constructing a Lagrangean description which interpolates between the NG and the Polyakov form. The interpolating theory thus offers a unified picture for understanding different features of the basic structures including their various symmetry properties. In this sense, therefore. it is more general than either the NG or Polyakov formulations. An added advantage is that it illuminates the passage from the NG form to the Polyakov form, which is otherwise lacking. In this context it may be noted that the Polyakov action has the additional Weyl invariance which NG action does not have. The interpolating action, which does not presuppose Weyl invariance, thus offers a proper platform of discussing the equivalence of the two actions. It also explains the emergence of the Weyl invariance in a natural way. The facilities of the interpolating Lagrangean formalism revealed in the string problem highlight the utility of the generalization of this formalism to the higher branes. We provide this generalization by taking the membrane as an example. Though our explicit construction refers to the membrane, most of the results will be applicable to the generic p - brane. The master Lagrangeans that interpolate between the different forms of the string and the membrane actions establish their equivalence from an alternative point of view. The reduction of the interpolating Lagrangean to the Polyakov form for the membrane is naturally plagued with the problem of the extra degrees of freedom of the membrane metric.As already stated, the emergence of the cosmological term is precisely from the accounting of these extra degrees of freedom. This formalism thus enables one to understand clearly the appearance of the cosmological term.

    Since the interpolating action formalism offers a composite scenario for discussing different features of such basic structures as strings, membranes etc, a thorough understanding of the gauge symmetries  occurring in this action is desirable, if not essential. We investigate this systematically using a Hamiltonian approach developed in the recent past \cite{brr}. The complete equivalence between the gauge and the diff. symmetries is established by providing an explicit map between the corresponding parameters. Both string and membrane models will be discussed separately. This is motivated by the fact that the interpolating action formalism for the string, though methodologically similar with the other branes, contains the above-mentioned unique features that demands separate treatment. Again, the results obtained from the string are useful for comparing results from the other models (p-branes) by going to the appropriate string limit.

   To put the work in a proper perspective, note that the role of symmetry has been recently emphasized in the loop quantum gravity quantization of string theory \cite{thi}. In particular, a construction has been discussed where the only local symmetry group is the diffeomorphism group and Weyl invariance is marginalized. We feel that the interpolating formulation fits into this general scheme because it is built on the gauge symmetry only. Recently the Arnowitt--Deser--Misner (ADM) representation used in general relativity has been utilized in the reduction of Polyakov Lagrangean for bosonic string and Green--Schwarz (GS) superstring \cite{deri}. The ADM representation uses the lapse and shift variables which owe their origin to the presence of first class constraints. The interpolating Lagrangean formalism, built as it is on the constraints which are implemented by Lagrange multipliers ( that are the analogues of the lapse and shift variables ), is seen to elucidate the connection with the ADM representation.

   The organization of the paper is as follows. In the next section we briefly review the interpolating action formalism for the string and its connection with the NG and the Polyakov theories. In Section-$3$ the constraint structure and gauge symmetry of the interpolating theory will be discussed for the string. Generalization of the above analysis to the membrane is contained in section-$4$. The concluding remarks will be presented in Section-$5$.

\section{The interpolating action of the string}

  The string is a one - dimensional object which will be assumed to be
embedded in the D - dimensional space-time. For simplicity we consider the embedding to be Minkowskian with mostly positive metric $\eta_{\mu\nu}$. The string
sweeps out a world - sheet which may be parametrized by two parameters $\tau$ and $\sigma$. The NG action for the free bosonic string is obtained from the integrated proper area of the world - sheet
\begin{eqnarray}
S_{\mathrm{NG}} = - \int d\tau d\sigma \left[\left(\dot{X}.X^{\prime}\right)^{2}
- \dot{X}^{2} X^{\prime 2}\right]^{\frac{1}{2}}
\label{111}
\end{eqnarray}
where $\dot{X}^{\mu} = \frac{\partial X^{\mu}}{ \partial\tau}$ and
${X}^{\prime \mu}
= \frac{\partial X^{\mu}}{\partial\sigma}$. 
The string tension is kept implicit for convenience. The string action is invariant under the reparametrization of the world - sheet i.e. under 
$\left( \tau,\sigma \right) \mapsto 
\left( \tau^{\prime}, \sigma^{\prime} \right)$
where
\begin{eqnarray}
\tau^{\prime} = \tau^{\prime}\left(\tau,\sigma\right)\nonumber\\
\sigma^{\prime} = \sigma^{\prime}\left(\tau,\sigma\right)
\label{112}
\end{eqnarray}
while the fields $X^{\mu}$ behave as scalars on the world - sheet,
\begin{equation}
X^{\prime\mu}\left(\tau^{\prime},\sigma^{\prime}\right) =
X^{\mu}\left(\tau,\sigma\right)
\label{113}
\end{equation}
The canonical momenta corresponding to the basic fields $X^{\mu}$  are
\begin{eqnarray}
\Pi_{\mu} = \frac{X^{\prime 2} \dot X_{\mu} - X^{\prime}_{\mu}
\left(\dot{X}.X^{\prime}\right)}{\left[\left(\dot{X}.X^{\prime}\right)^{2} -
  \dot{X}^{2} X^{\prime 2}\right]^{\frac{1}{2}}}
\label{114}
\end{eqnarray}
     From the definition (\ref{114}) we get the primary constraints
for the NG string
\begin{eqnarray}
\Omega_{1} = \Pi_{\mu}X^{\prime \mu} \approx 0 \nonumber\\
\Omega_{2} = \Pi^{2} + X^{\prime 2} \approx 0
\label{115}
\end{eqnarray}
The nontrivial Poisson's brackets of the theory are given by
\begin{eqnarray}
 \{X^{\mu}\left(\tau,\sigma\right),
 \Pi_{\nu}\left(\tau,\sigma^{\prime}\right)\} = \eta_{\nu}^{\mu}
 \delta\left(\sigma - \sigma^{\prime}\right)
\label{116}
\end{eqnarray}
Using these Poisson brackets it is easy to work out the algebra of the
constraints
\begin{equation}
\left\{ \Omega_{1}\left(\sigma\right),\Omega_{1}\left(\sigma^{\prime}\right)
\right\} = \left(\Omega_{1}\left(\sigma\right) +
\Omega_{1}\left(\sigma^{\prime}\right)\right)\partial_{\sigma}
\delta\left(\sigma - \sigma^{\prime}\right)
\label{117}
\end{equation}
\begin{equation}
\left\{ \Omega_{1}\left(\sigma\right),\Omega_{2}\left(\sigma^{\prime}\right)
\right\} = \left(\Omega_{2}\left(\sigma\right) +
\Omega_{2}\left(\sigma^{\prime}\right)\right)\partial_{\sigma}
\delta\left(\sigma - \sigma^{\prime}\right)
\label{118}
\end{equation}
\begin{equation}
\left\{ \Omega_{2}\left(\sigma\right),\Omega_{2}\left(\sigma^{\prime}\right)
\right\} = 4 \left(\Omega_{1}\left(\sigma\right) +
\Omega_{1}\left(\sigma^{\prime}\right)\right)
\partial_{\sigma}\delta\left(\sigma - \sigma^{\prime}\right)
\label{119}
\end{equation}

   Clearly, the Poisson brackets between the constraints (\ref{115}) are weakly involutive so that the set (\ref{115}) is first class. The canonical Hamiltonian is
\begin{eqnarray}
H_{c} = \Pi_{\mu}\dot X^{\mu} - {\cal{L}}
\label{1110}
\end{eqnarray}
Substituting the appropriate expressions in (\ref{1110}) we observe that the
canonical Hamiltonian vanishes, as expected for a reparametrization
invariant theory. The total Hamiltonian is thus expressed as a linear combination of the first-class constraints (\ref{115}),
\begin{equation}
H_{T} = - \rho \Omega_{1} - \frac{\lambda}{2}\Omega_{2}
\label{1111}
\end{equation}
where $\rho$ and $\lambda$ are Lagrange multipliers. Conserving the primary constraints no  new secondary constraints emerge. The total set of constraints of the NG theory is then given by the first class system (\ref{115}).

  After reviewing the salient features of the NG action of the string we now pass on to the construction of the interpolating action \cite{rbs}. To achieve this end we write the Lagrangean of the NG action in the first order form \cite{ka}.
\begin{equation}
{\cal{L}}_{I} = \Pi_{\mu} \dot X^{\mu} - H_{T}
\label{1113}
\end{equation}
Substituting $H_{T}$ from (\ref{1111}), ${\cal{L}}_{I}$ becomes
\begin{equation}
{\cal{L}}_{I} = \Pi_{\mu}\dot X^{\mu} + \rho \Pi_{\mu}X^{\prime \mu}
 + \frac{\lambda}{2}\left(\Pi^{2} + X^{\prime 2} \right)
\label{1114}
\end{equation}
In the above equation $\lambda$ and $\rho$, originally introduced
as lagrange multipliers, will be treated as independent
fields. Since $\Pi_{\mu}$ is really an auxiliary field we will eliminate
it from (\ref{1114}). The Euler-Lagrange equation for $\Pi_{\mu}$ is
\begin{equation}
\dot X^{\mu}+ \rho X^{\prime \mu} + \lambda \Pi^{\mu}  = 0
\label{1115}
\end{equation}
Substituting $\Pi_{\mu}$ back in (\ref{1114}) the Lagrangian ${\cal{L}}_{I}$
reduces to
\begin{equation}
{\cal{L}}_{I} = -\frac{1}{2 \lambda}\left[\dot X^{2} +
2 \rho \dot X_{\mu}X^{\prime \mu} + \left( \rho^{2} - \lambda^{2}\right)
X^{\prime 2}\right]
\label{1116}
\end{equation}
We will call the Lagrangean (\ref{1116}) as the Interpolating Lagrangean of the
bosonic string. The justification of the name will be established below by showing that passing to the appropriate limits one can derive the NG action and the Polyakov action from (\ref{1116}) \cite{rbs}.

    The reproduction of the NG action from the interpolating action is trivial. To pass to this limit we need only to eliminate the extra fields $\rho$ and
$\lambda$ from (\ref{1116}).
The solution of the E-L equation for $\rho$ and $\lambda$ following from
(\ref{1116}) are
\begin{equation}
\rho = - \frac{\dot X^{\mu} X^{\prime}_{\mu}}{X^{\prime 2}}
\label{1117}
\end{equation}
and
\begin{equation}
\lambda^{2} = \frac{\left(\dot{X}.X^{\prime}\right)^{2}
- \dot{X}^{2} X^{\prime 2}}{X^{\prime 2} X^{\prime 2}}
\label{1118b}
\end{equation}
From (\ref{1118b}) $\lambda$ is determined modulo a sign which can be fixed by demanding the consistency of (\ref{114}) with (\ref{1115}). Accordingly
\begin{equation}
\lambda = -\frac{\left[\left(\dot{X}.X^{\prime}\right)^{2}
- \dot{X}^{2} X^{\prime 2}\right]^{\frac{1}{2}}}{X^{\prime 2}}
\label{1118}
\end{equation}
Now, substituting $\rho$ and $\lambda$ from (\ref{1117}) and (\ref{1118}) in ${\cal{L}}_{I}$ we get back the NG form.

 The reproduction of the Polyakov action from (\ref{1116}) is not so straightforward. The Polyakov action for the free bosonic string is given by
\begin{equation}
S_{\mathrm{P}} = -\frac{1}{2}\int d^2\xi {\sqrt - g} g^{ij} \partial_{i} X^{\mu}
\partial_{j} X_{\mu}
\label{1112}
\end{equation}
Here $g_{ij}$ are considered as independent fields while $\xi^i$ collectively denote the parameters, $\xi^0 = \tau$ and $\xi^1 = \sigma$. The merit of (\ref{1112})
in the path integral quantization was first pointed out by Polyakov. Compared with the NG version, the action (\ref{1112}) contains additional fields $g_{ij}$. It is equivalent to the NG action in the sense that solving $ g_{ij} $ in (\ref{1112}) from its equation of motion one can reproduce the NG action.
In the following we will get a better understanding of the correspondence through the interpolating action. The Polyakov action is a more redundant description of the string than the NG form because of the extra fields $g_{ij}$ introduced in it. The reparametrization invariance under (\ref{112}) is ensured by the transformations (\ref{113}) along with the transformations
\begin{equation}
g^{\prime}_{ij}(\xi^{\prime}) = \frac{\partial\xi^{k}}{\partial\xi^{\prime i}}
                   \frac{\partial\xi^{l}}{\partial\xi^{\prime j}} g_{kl}(\xi)
\label{113A}
\end{equation}
 Looking at the transformation relations under (\ref{113}) and (\ref{113A}) it is apparent that the reparametrization invariance is synonymous with general covariance on the world - sheet with $X^{\mu}$ transforming as scalar fields. Apart from the reparametrization invariance the Polyakov string has the Weyl invariance
\begin{equation}
g^{\prime}_{ij}(\xi) = \exp\left(\Lambda\left(\xi\right)\right)g_{ij}(\xi)
\label{113B}
\end{equation}
where $\Lambda (\xi)$ is any arbitrary well behaved function of $\xi$. Though there are three different metric coefficients $g_{ij}$, due to the existence of this scale (Weyl) invariance only two of them are really independent. The Weyl invariance is special to the Polyakov string, the higher branes do not share it. Clearly, in the Polyakov action of the string there are two independent fields only apart from $X_{\mu}$. These are the two independent components of the metric. These two components can also be fixed by the two reparametrization symmetries. Usually the light-cone metric ${\rm {diag}}(1, -1)$ is employed in the gauge fixed calculation. However here we work in the gauge independent approach, otherwise the interplay of gauge and diff. symmetries is lost. That the metric is completely determinable is manifested in our approach by the exact matching of the number of independent metric components with the number of extra fields in the interpolating Lagrangean (\ref{1116}). It will thus be possible to map the interpolating Lagrangean to the Polyakov form in a unique manner. We take the following Ansatz \cite{rbs}\footnote {Such a representation was also discussed by Giddings \cite{gid}}
\begin{equation}
g^{ij} = \left(-g\right)^{-\frac{1}{2}}
\left(\begin{array}{cc}
\frac{1}{\lambda}&\frac{\rho}{\lambda}\\
\frac{\rho}{\lambda}&\frac{\rho^2 - \lambda^2}{\lambda}
\end{array}\right)
\label{1119}
\end{equation}
where $g = {\rm{det}}g_{ij}$ and $g^{ij}$ is the inverse of $g_{ij}$.
With this choice the Interpolating Lagrangean
reduces to
\begin{equation}
{\cal{L}}_{I} = -\frac{1}{2}{\sqrt - g} g^{ij} \partial_{i} X^{\mu}
\partial_{j} X_{\mu}
\label{1120}
\end{equation}
Clearly this Lagrangean corresponds to the Polyakov form of the string action. We next verify the consistency of the construction (\ref{1119}). As a first step, note that,
\begin{equation}
{\rm det}g^{ij} = \left( - g\right)^{-1}\left[ \frac{\rho^{2} -
\lambda^{2}}{\lambda^{2}} - \frac{\rho^{2}}{\lambda^{2}}\right] = \frac{1}{g}
\label{1121}
\end{equation}
as it should be because $g^{ij}$ is the inverse matrix of $g_{ij}$. Further,
from the identification (\ref{1119}) we find
\begin{equation}
g^{00} = \left( - g \right)^{-\frac{1}{2}}\frac{1}{\lambda}
\label{1122}
\end{equation}
and
\begin{equation}
g^{01} = g^{10} = \left( - g \right)^{-\frac{1}{2}}\frac{\rho}{\lambda}
\label{1123}
\end{equation}
From the above equations we can solve $\rho$ and $\lambda$ in terms of
$g^{00}$ and $g^{01}$ as
\begin{eqnarray}
\rho &=&\frac{g^{01}}{g^{00}}\nonumber\\
\lambda &=& \frac{1}{\left( \sqrt{- g }\right)\left( g^{00}\right)}
\label{1124}
\end{eqnarray}
Finally the mapping (\ref{1119}) also yields
\begin{equation}
g^{11} = \frac{\rho^{2} -\lambda^{2}}{\lambda}
\left( - g \right)^{-\frac{1}{2}}
\label{1125}
\end{equation}
After substituting the solutions for $\rho$ and $\lambda$, the resulting expression becomes
\begin{equation}
\left(g^{01} \right)^{2} -  \left(\sqrt{ - g } \right)^{-2} = g^{11} g^{00}
\label{1126}
\end{equation}
which, after a simple rearrangement, is shown to be the same as the condition (\ref{1121}). This completes the consistency check of the construction (\ref{1119}). 

 Before concluding this section, we note that the interpolating Lagrangean formalism introduces a metric $g_{ij}$ on the string world sheet, the elements of which are constructed from the fields $\rho$ and $\lambda$ of the interpolating Lagrangean. These are the Lagrange multipliers which enforce the first class constraints of the theory. Our construction is reminiscent of the Arnowitt--Deser--Misner (ADM) representation \cite{mtw}\footnote{We thank A.~A.~Deriglazov for pointing this out.} used in general relativity. In the ADM representation the metric of the four dimensional Riemannian space time ${}^{\left(4\right)}\gamma^{\mu\nu}$ is split as\footnote{For the metric of the total space time the dimension is mensioned as a (pre)superscript} 
\begin{equation}
\left(\begin{array}{cc}
{}^{\left(4\right)}\gamma^{00}&{}^{\left(4\right)}\gamma^{0m}\\
{}^{\left(4\right)}\gamma^{0k}&{}^{\left(4\right)}\gamma^{km}
\end{array}\right)
 =\left(\begin{array}{cc}
-\frac{1}{\left(N\right)^{2}}&\frac{\left(N^{m}\right)}{\left(N\right)^{2}}\\
\frac{\left(N^{k}\right)}{\left(N\right)^{2}}&\left(\gamma^{km}-\frac{\left(N^{k}\right)\left(N^{m}\right)}{\left(N\right)^{2}}\right)
\end{array}\right)
\label{ADM4}
\end{equation}
where, $k$, $m$ take the values $1$, $2$, $3$. $\gamma^{km}$ is the metric on a three dimensional hypersurface embedded in the four dimensional space time. $N$ and $N^{k}$ are the arbitrary lapse and shift veriables which are nothing but the Lagrange multipliers of the theory. A similar representation for $d = 2$ assumes the following form 
\begin{equation}
\left(\begin{array}{cc}
{}^{\left(2\right)}\gamma^{00}&{}^{\left(2\right)}\gamma^{01}\\
{}^{\left(2\right)}\gamma^{01}&{}^{\left(2\right)}\gamma^{11}
\end{array}\right)
 =\left(\begin{array}{cc}
-\frac{1}{\left(N\right)^{2}}&\frac{\left(N^{1}\right)}{\left(N\right)^{2}}\\
\frac{\left(N^{1}\right)}{\left(N\right)^{2}}&\left(\gamma^{11}-\frac{\left(N^{1}\right)^{2}}{\left(N\right)^{2}}\right)
\end{array}\right)
\label{ADM2}
\end{equation}
From (\ref{ADM2}) we can easily calculate 
\begin{eqnarray}
\mathrm{det} \gamma^{ij} &=& - \gamma^{11}/\left(N\right)^{2} \nonumber \\
\mathrm{i.e.}\quad \left(\gamma\right)^{-1} &=& -\gamma^{11}/\left(N\right)^{2}
\label{ADM1}
\end{eqnarray}
where $\gamma = \mathrm{det}\gamma_{ij}$.
Now if we introduce the correspondence 
\begin{eqnarray}
N^{1}\mapsto -\rho \quad \mathrm {and}\quad \left(N\right)^{2}\mapsto -\lambda\sqrt{-g}\quad \mathrm{and}\quad \gamma\mapsto g
\label{ADM3}
\end{eqnarray}
and use (\ref{ADM1}), (\ref{ADM2}) becomes identical with our construction (\ref{1119}). Recently, the ADM representation (\ref{ADM2}) was used to express Lagrangeans for Bosonic strings as well as Green--Schwarz (GS) superstring, starting from the Polyakov Lagrangeans \cite{deri}. The new feature contained in our interpolating Lagrangean is the precise mechanism by which an independent metric is introduced on the string world sheet. This formalism also exemplifies the correspondence between the ADM (lapse and shift) variables with the Lagrange multipliers in the context of string. In this connection it may be noted that $\gamma^{11}$ is expressed by (\ref{ADM1}) in terms of $N$ with the choice of a scale $\left(\gamma^{-1}\right)$. This is related with the presence of Weyl invariance for the string. Later, we will observe that in the example of the membrane, there is no such freedom. For the membrane (or the higher brane), the metric on the hypersurface must be considered as completely arbitrary. We will also observe a similar correspondence like (\ref{ADM3}) for the example of the membrane.

\section{Constraint structure and gauge symmetry}

     In this section we will discuss the gauge symmetries of the different actions and find their exact correspondence with the reparametrization invariances. Since our discussion will be centered on the interpolating action (\ref{1116}) let us begin with an analysis of its constraint structure. The independent fields in (\ref{1116}) are $X_{\mu}$, $\rho$ and $\lambda$. Let the corresponding momenta be denoted by $\Pi_{\mu}$, $\Pi_{\rho}$ and $\Pi_{\lambda}$ respectively. By definition
\begin{eqnarray}
\Pi_{\mu}& =&  -\frac{1}{\lambda}\left(\dot{X}_{\mu} +
               \rho X^{\prime}_{\mu}\right)\nonumber\\
\Pi{\rho}& = & 0 \nonumber\\
\Pi_{\lambda}& = & 0
\label{211}
\end{eqnarray}
In addition to the Poisson brackets similar to (\ref{116}) we now have
\begin{eqnarray}
 \{\rho\left(\tau,\sigma\right),
 \Pi_{\rho}\left(\tau,\sigma^{\prime}\right)\} =
  \delta\left(\sigma - \sigma^{\prime}\right) \nonumber\\
 \{\lambda\left(\tau,\sigma\right),
 \Pi_{\lambda}\left(\tau,\sigma^{\prime}\right)\} =
  \delta\left(\sigma - \sigma^{\prime}\right)
\label{212}
\end{eqnarray}
The canonical Hamiltonian following from (\ref{1116}) is
\begin{equation}
{\cal{H}}_c = -\rho\Pi_{\mu}X^{\prime\mu} 
              - \frac{\lambda}{2}\left(\Pi^2 +X^{\prime 2}\right) 
\label{213}
\end{equation}
which reproduces the total Hamiltonian of the NG action. From the
definition
of the canonical momenta we can easily identify the primary constraints
\begin{eqnarray}
\Pi{\rho}& \approx & 0 \nonumber\\
\Pi_{\lambda}& \approx & 0
\label{214}
\end{eqnarray}
Conserving these primary constraints we find that two new secondary
constraints emerge. These are the constraints
$\Omega_{1}$ and $\Omega_{2}$ of equation (\ref{115}), as expected.
The primary constraints of the NG action
appear as secondary constraints in this formalism. No more secondary
constraints are obtained. The system of constraints for the Interpolating
Lagrangean thus comprises of the set (\ref{115}) and (\ref{214}). The Poisson
brackets of the constraints of (\ref{214}) vanish within themselves. Also
the PB of these with (\ref{115}) vanish. All the constraints are first class
and therefore generate gauge transformations on ${\cal{L}}_I$ but the number
of independent gauge parameters is equal to the number of independent primary
first class constraints i.e. two.  In the following analysis we will apply
a systematic procedure of abstracting
the most general local symmetry transformations of the Lagrangean. A brief
review of the procedure of \cite{brr} will thus be appropriate.

    Consider a theory with first class constraints only. The set of
constraints $\Omega_{a}$ is assumed to be classified as
\begin{equation}
\left[\Omega_{a}\right] = \left[\Omega_{a_1}
                ;\Omega_{a_2}\right]
\label{215}
\end{equation}
where $a_1$ belong to the set of primary and $a_2$ to the set of
secondary constraints. The total Hamiltonian is
\begin{equation}
H_{T} = H_{c} + \Sigma\lambda^{a_1}\Omega_{a_1}
\label{216}
\end{equation}
where $H_c$ is the canonical Hamiltonian and $\lambda^{a_1}$ are Lagrange multipliers enforcing the primary constraints. The most general expression for the generator of gauge transformations is obtained according to the Dirac conjecture as
\begin{equation}
G = \Sigma \epsilon^{a}\Omega_{a}
\label{217}
\end{equation}
where $\epsilon^{a}$ are the gauge parameters, only $a_1$ of which are independent. By demanding the commutation of an arbitrary gauge variation with the total time derivative,(i.e. $\frac{d}{dt}\left(\delta q \right) = \delta \left(\frac{d}{dt} q \right) $) we arrive at the following equations \cite{brr,htz}
\begin{equation}
\delta\lambda^{a_1} = \frac{d\epsilon^{a_1}}{dt}
                 -\epsilon^{a}\left(V_{a}^{a_1}
                 +\lambda^{b_1}C_{b_1a}^{a_1}\right)
                              \label{218}
\end{equation}
\begin{equation}
  0 = \frac{d\epsilon^{a_2}}{dt}
 -\epsilon^{a}\left(V_{a}^{a_2}
+\lambda^{b_1}C_{b_1a}^{a_2}\right)
\label{219}
\end{equation}
Here the coefficients $V_{a}^{a_{1}}$ and $C_{b_1a}^{a_1}$ are the structure
functions of the involutive algebra, defined as
\begin{eqnarray}
\{H_c,\Omega_{a}\} = V_{a}^b\Omega_{b}\nonumber\\
\{\Omega_{a},\Omega_{b}\} = C_{ab}^{c}\Omega_{c}
\label{2110}
\end{eqnarray}
Solving (\ref{219}) it is possible to choose $a_1$ independent
gauge parameters from the set $\epsilon^{a}$ and express $G$ of
(\ref{217}) entirely in terms of them. The other set (\ref{218})
gives the gauge variations of the Lagrange multipliers. It can be shown that
these equations are not independent conditions but appear as internal
consistency conditions. In fact the conditions (\ref{218}) follow from
(\ref{219}) \cite{brr}.

We begin the analysis with the NG action (\ref{111}).
Here the only fields are $X^{\mu}$.
The generator of the gauge transformations of (\ref{111})
is obtained from the constraints
$\Omega_{i}$ given by (\ref{115}) as
\begin{equation}
G =\int d\sigma \alpha_{i}\Omega_{i}
\label{3112}
\end{equation}
where $\alpha_{i}$ are the independent gauge parameters.
The transformations of $X^{\mu}$  under (\ref{3112})
can be worked out resulting in the following
\begin{equation}
\delta X_{\mu} = \left\{X_{\mu}, G \right\}
 = \left( \alpha_{1} X_{\mu}^{\prime} + 2 \alpha_{2} \Pi_{\mu} \right)
\label{3114}
\end{equation}
Substituting $\Pi_{\mu}$ from (\ref{114}) in the above we get the appropriate
gauge transformation of $X_{\mu}$ that leave (\ref{111}) invariant,
\begin{equation}
\delta X_{\mu}
 = \left(\alpha_1 - \frac{2\alpha_2\left(\dot{X}.X^{\prime}\right)}
{\left[\left(\dot{X}.X^{\prime}\right)^{2} -
  \dot{X}^{2} X^{\prime 2}\right]^{\frac{1}{2}}}
\right)X^{\prime}_{\mu} + \frac{ 2 \alpha_{2}X^{\prime 2}}
{\left[\left(\dot{X}.X^{\prime}\right)^{2} -
  \dot{X}^{2} X^{\prime 2}\right]^{\frac{1}{2}}}
\dot{X_{\mu}}
\label{31141}
\end{equation}
Identifying the coefficients of $\dot{X}$ and $X^{\prime}$ respectively with
$\Lambda_{0}$ and $\Lambda_{1}$ we get
\begin{equation}
\delta X_{\mu}
 = \Lambda_{0} \dot{X_{\mu}} + \Lambda_{1} X_{\prime}{\mu}
\label{31142}
\end{equation}
Note that $\Lambda_{0}$ and $\Lambda_{1}$ are arbitrary functions of the
parameters $\xi_i$. Using (\ref{113}) we observe that these gauge variations
(\ref{31142}) coincide with the variations due to the reparametrization
\begin{eqnarray}
\tau^{\prime}& = & \tau - \Lambda_{0}\nonumber\\
\sigma^{\prime}& = & \sigma - \Lambda_{1}
\label{31143}
\end{eqnarray}
The complete mapping of the gauge transformations with the
reparametrizations is thus established for the NG string.

   We then take up the interpolating Lagrangean (\ref{1116}). It contains additional fields $\rho$ and $\lambda$. The full constraint structure of the theory comprises of the constraints (\ref{214}) along with (\ref{115}). We could proceed from these and construct the generator of gauge transformations from (\ref{3112}) by including the whole set of first class constraints. Using (\ref{219}) the dependent gauge parameters can be eliminated. After finding the gauge generator in terms of the independent gauge parameters, the variations of the fields $X^{\mu}$, $\rho$ and $\lambda$ can be worked out. However, looking at the intermediate first order form (\ref{1115}) we understand that the variations of the fields $\rho$ and $\lambda$ can be calculated alternatively, ( using (\ref{218})) from the NG theory where they appear as Lagrange multipliers. We adopt this alternative procedure. The generator of gauge transformations has already been given in (\ref{3112}). So the gauge variations of $X^{\mu}$ is again given by (\ref{3114}). Next, we relabel $\rho$ and $\lambda$ by $\lambda_{1}$ and $\lambda_{2}$, where
\begin{equation}
\lambda_{1} = \rho \hspace{1cm} \rm{and} \hspace{1cm}
\lambda_{2} = \frac{\lambda}{2}
\label{313}
\end{equation}
The  variations of $ \lambda_{i}$ are obtained from (\ref{218})
\begin{equation}
\delta\lambda_{i}\left( \sigma\right) = -  \dot \alpha_{i}
- \int d\sigma^{\prime} d\sigma^{\prime \prime} C_{kj}{}^{i}
\left(\sigma^{\prime}, \sigma^{\prime \prime}, \sigma\right)
\lambda_{k}\left(\sigma^{\prime}\right) \alpha_{j}
\left(\sigma^{\prime \prime}\right)
\label{314}
\end{equation}
where
$ C_{kj}{}^{i}
\left(\sigma^{\prime}, \sigma^{\prime \prime}, \sigma\right)$
are given by
\begin{equation}
\left\{ \Omega_{\alpha}\left(\sigma\right),\Omega_{\beta}
\left(\sigma^{\prime}\right)\right\} =  \int d\sigma^{\prime \prime}
C_{\alpha \beta}{}^{\gamma}\left(\sigma, \sigma^{\prime}, \sigma^{\prime
\prime}\right) \Omega_{\gamma}\left(\sigma^{\prime \prime}\right)
\label{315}
\end{equation}
Observe that the structure function $ V_{a}{}^{b}$ does not appear in (\ref{314}) since $H_{c} = 0$ for the NG theory. The nontrivial structure functions $C_{\alpha \beta}{}^{\gamma}\left(\sigma, \sigma^{\prime}, \sigma^{\prime \prime}\right)$ are obtained from the
constraint algebra (\ref{117} - \ref{119}) as
\begin{equation}
C_{1 1}{}^{1}\left(\sigma ,\sigma^{\prime},\sigma^{\prime \prime}\right) =
\partial_{\sigma}\delta\left(\sigma - \sigma^{\prime}\right)
\left(\delta\left(\sigma - \sigma^{\prime \prime}\right) +
\delta\left(\sigma^{\prime} - \sigma^{\prime \prime}\right) \right)
\label{316}
\end{equation}
\begin{equation}
C_{1 2}{}^{2}\left(\sigma, \sigma^{\prime}, \sigma^{\prime \prime}\right) =
\partial_{\sigma}\delta\left(\sigma - \sigma^{\prime}\right)
\left(\delta\left(\sigma - \sigma^{\prime \prime}\right) +
\delta\left(\sigma^{\prime} - \sigma^{\prime \prime}\right) \right)
\label{317}
\end{equation}
\begin{equation}
C_{2 1}{}^{2}\left(\sigma, \sigma^{\prime}, \sigma^{\prime \prime}\right) =
\partial_{\sigma}\delta\left(\sigma - \sigma^{\prime}\right)
\left(\delta\left(\sigma - \sigma^{\prime \prime}\right) +
\delta\left(\sigma^{\prime} - \sigma^{\prime \prime}\right) \right)
\label{318}
\end{equation}
\begin{equation}
C_{2 2}{}^{1}\left(\sigma, \sigma^{\prime}, \sigma^{\prime \prime}\right) =
4 \partial_{\sigma}\delta\left(\sigma - \sigma^{\prime}\right)
\left(\delta\left(\sigma - \sigma^{\prime \prime}\right) +
\delta\left(\sigma^{\prime} - \sigma^{\prime \prime}\right) \right)
\label{319}
\end{equation}
all other $ C_{\alpha b}{}^{\gamma}$'s are zero.
Using the expressions of the structure functions (\ref{316} - \ref{319})
in equation(\ref{314}) we can easily derive
\begin{eqnarray}
\delta \lambda_{1} &=& - \dot \alpha_{1}
+ \left(\alpha_{1}\partial_{1}\lambda_{1}
 - \lambda_{1}\partial_{1}\alpha_{1} \right) 
+ 4 \left(\alpha_{2}\partial_{1}\lambda_{2} - \lambda_{2}\partial_{1}
\alpha_{2}\right)\nonumber\\
\delta \lambda_{2} &=& -\dot \alpha_{2}
+\left(\alpha_{2}\partial_{1}\lambda_{1} - \lambda_{1} \partial_{1}\alpha_{2}
\right)
+\left(\alpha_{1}\partial_{1}\lambda_{2} - \lambda_{2} \partial_{1}\alpha_{1}
\right)
\label{GM41}
\end{eqnarray}
From the correspondence (\ref{313}) we get the variations of $\rho$
and $\lambda$ as
\begin{eqnarray}
\delta \rho &=& - \dot \alpha_{1}
+ \left(\alpha_{1}\partial_{1}\rho
 - \rho\partial_{1}\alpha_{1} \right) + 2 \left(\alpha_{2}\partial_{1}\lambda - \lambda \partial_{1}\alpha_{2}\right)
\nonumber\\
\delta \lambda &=& -2 \dot \alpha_{2}
+2 \left(\alpha_{2}\partial_{1}\rho - \rho \partial_{1}\alpha_{2}
\right)
+\left(\alpha_{1}\partial_{1}\lambda - \lambda \partial_{1}\alpha_{1}\right)
\label{GM51}
\end{eqnarray}
Note that the above expressions for the gauge variations of $\rho$ and $\lambda$ can also be obtained from their definitions (\ref{1117}), (\ref{1118}) and the expression of gauge variation of $X^{\mu}$ (\ref{31141}) . We get from equation (\ref{1117})
\begin{eqnarray}
\delta \rho & = & - \delta \left( \frac{\partial_{0} X^{\mu}\partial_{1} X_{\mu}}{\partial_{1} X^{\mu}\partial_{1} X_{\mu}}\right)\nonumber\\
            & = & - \left\{ \frac{\partial_{1} X_{\nu}}{\partial_{1} X^{\mu}\partial_{1} X_{\mu}}\right\}\partial_{0}\left(\delta X^{\nu}\right) 
+ \left\{2\frac{\partial_{0} X^{\mu}\partial_{1} X_{\mu} \partial_{1} X_{\nu}}
{\left(\partial_{1} X^{\mu}\partial_{1} X_{\mu}\right)^{2}} - \frac{\partial_{0} X_{\nu}}{\partial_{1} X^{\mu}\partial_{1} X_{\mu}} \right\}\partial_{1}\left(\delta X^{\nu}\right)\nonumber\\
\label{GM61}
\end{eqnarray}
which relates the gauge variation of $\rho$ with that of $X^{\mu}$. Using the definitions (\ref{1117}) and (\ref{1118}) the gauge variation of $X^{\mu}$ given by (\ref{31141}) can be reduced to the following convenient form 
\begin{eqnarray}
\delta X^{\mu} = \left( \alpha_{1} - 2 \frac{\alpha_{2}\rho}{\lambda}\right)\partial_{1}X^{\mu} - \frac{2 \alpha_{2}}{\lambda}\partial_{0} X^{\mu}
\label{GM71}
\end{eqnarray}
Substituting $\delta X^{\mu}$ from (\ref{GM71}) in (\ref{GM61}) we recover, after some simplification, the same expression for $\delta \rho$ as in (\ref{GM51}). Similarly, $\delta \lambda $ can also be computed directly from the definition of $\lambda$. We first note that $\rho$ and $\lambda$ can be related as
\begin{eqnarray}
\rho^{2} - \lambda^{2} = \left( \frac{\partial_{0} X^{\mu}\partial_{0} X_{\mu}}{\partial_{1} X^{\mu}\partial_{1} X_{\mu}}\right)
\label{GM81}
\end{eqnarray}
which follows from equation (\ref{1117}) and (\ref{1118}). From the relation (\ref{GM81}) we can easily derive that 
\begin{eqnarray}
2 \rho \delta \rho - 2 \lambda \delta \lambda  =  \left\{ \frac{2 \partial_{0} X_{\nu}}{\partial_{1} X^{\mu}\partial_{1} X_{\mu}}\right\}\partial_{0}\left(\delta X^{\nu}\right) - \left\{ \frac{2 \partial_{0} X^{\mu}\partial_{0} X_{\mu}\partial_{1} X^{\nu}}{\left(\partial_{1} X^{\mu}\partial_{1} X_{\mu}\right)^{2}}\right\}\partial_{1}\left(\delta X^{\nu}\right)
\label{GM91}
\end{eqnarray}
Equation (\ref{GM91}) enables us to find $\delta \lambda$ from the known expressions of $\delta X^{\mu}$ and $\delta \rho$. The resulting expression of $\delta \lambda$ is identical with that given in (\ref{GM51}). This observation again confirms our remark about (\ref{218}) that those are really internal consistency conditions \cite{brr}.

     In the above we have found out the full set of symmetry transformations
of the fields in the interpolating Lagrangean (\ref{1116}). Clearly, the
same set of transformations apply to the first order form (\ref{1115}).
In the latter, $\Pi_{\mu}$ is introduced as an additional field. Its appropriate gauge transformation is not difficult to find
\begin{equation}
\delta \Pi_{\mu} = \left\{\Pi_{\mu}, G \right\}
 = \left( \Pi_{\mu} \alpha_{1} + 2 X_{\mu}^{\prime}\alpha_{2} \right)^{\prime}
\label{3113}
\end{equation}
The symmetry transformations (\ref{GM51}) were earlier given in \cite{ka}. But the results were found there by inspection\footnote
{ For easy comparison identify
$ \alpha_{1} = \eta $ and $ 2 \alpha_{2} = \epsilon $ }.
In our approach the appropriate transformations are
obtained systematically by a general method applicable to a whole class of
actions.

     At this stage we concentrate our attention on the Polyakov action (\ref{1112}). We can take it as an independent example for the application of our analysis based on (\ref{219}). In the Polyakov version the set of basic fields contain $g_{ij}$ apart from the fields $X^{\mu}$. Working out the full set of constraints we can construct the gauge generator $G$ according to (\ref{3112}). The set of constraints $\Omega_{i}$ include all the first class constraints, both primary and secondary. We then have to invoke (\ref{219}) to solve the dependent gauge parameters. These solutions, when substituted in the expression of $G$ give the desired form of $G$ in terms of the independent number of gauge parameters. The gauge variation of $g_{ij}$ can then be computed by the usual procedure
\begin{equation}
\delta g_{ij} = \left\{ g_{ij}, G \right\}
\label{hati}
\end{equation}
However, a particular usefulness of the interpolating Lagrangean formalism can be appreciated now. It is not required to find the gauge variations of $g_{ij}$ from scratch. The identification (\ref{1119}) allows us to find the required gauge variations from the corresponding transformations of $\rho$ and $\lambda$. This possibility is actually a consequence of the essential unity of the nature of gauge symmetries in different versions of the string actions. This underlying unity is the reparametrization invariance of the actions which is manifested in the form of gauge symmetries \cite{jvh}. We have already indicated this for the NG model. The Polyakov action offers a more important platform to test this proposition. Indeed, the complete equivalence between the two concepts can be demonstrated from the Polyakov action by devising an exact mapping between the reparametrization parameters and gauge parameters by comparing the changes of $\rho$ and $\lambda$ from the alternative approaches using the identification (\ref{1119}).
 
     We have already noted how the fields $X_{\mu}$ and $g_{ij}$ behave under
reparametrization (see equations (\ref{113}) and (\ref{113A})). Considering
infinitesimal transformation (\ref{31143})
we can write the variation of $X_{\mu}$ as
\begin{equation}
\delta X^{\mu} = \Lambda^{i} \partial_{i} X^{\mu}
= \Lambda^{0} \dot X^{\mu} + \Lambda^{1}  X^{\prime \mu}
\label{3117}
\end{equation}
and that of $g_{ij}$
\begin{equation}
\delta g_{ij} = D_{i}\Lambda_{j} + D_{j}\Lambda_{i}
\label{3122}
\end{equation}
where
\begin{equation}
D_{i}\Lambda_{j} = \partial_{i} \Lambda_{j} - \Gamma_{ij}{}^{k} \Lambda_{k}
\label{3123}
\end{equation}
$\Gamma_{ij}{}^{k}$ being the usual Christoffel symbols \cite{jvh}. 

      The infinitesimal parameters $\Lambda^{i}$ characterizing reparametrization correspond to infinitesimal gauge transformations and will ultimately be related with the gauge parameters $\alpha_{i}$ introduced earlier such that the symmetry transformations on $X^{\mu}$ agree from both the approaches. Since the metric $g_{ij}$ is associated with $\rho$ and $\lambda$ by the correspondence (\ref{1119}), equation (\ref{3122}) will enable us to establish the desired mapping between the gauge and the reparametrization parameters.
    
     To compare variations of $X^{\mu}$ from the alternative approaches we proceed as follows. From the Lagrangean corresponding to (\ref{1112}) we find
\begin{equation}
\Pi^{\mu} = - \sqrt{- g } g^{00} \dot X^{\mu} - \sqrt{- g } g^{01}
 X^{\prime \mu}
\label{3118}
\end{equation}
Substituting $\dot X^{\mu}$ from (\ref{3118}) in (\ref{3117}) we get after
 some calculation
\begin{equation}
\delta X^{\mu} = \Lambda^{0} \frac{\sqrt{- g } }{g g^{00} }
 \Pi^{\mu}  + X^{\mu \prime}\left(\Lambda^{1} - \frac{g^{01} }{g^{00} }
 \Lambda^{0}\right)
\label{3119}
\end{equation}
Comparing the above expression of $\delta X^{\mu}$ with that of (\ref{3114})
we find the mapping
\begin{eqnarray}
\Lambda^{0} = -2 \sqrt{- g } g^{00} \alpha_{2}    =
- \frac{ 2 \alpha_{2}}{\lambda}
\nonumber \\
\Lambda^{1} = \alpha_{1} - 2 \sqrt{- g } g^{01} \alpha_{2}    = \alpha_{1} -
\frac{ 2 \rho \alpha_{2}}{\lambda}
\label{3120}
\end{eqnarray}
With this mapping the gauge transformation on $X_{\mu}$ in both the formalism
agree.

     We will now compute $\delta \rho$ and $\delta \lambda$ from $\delta g_{ij}$ given by (\ref{3122}) using the identification (\ref{1119}). With the help of the mapping (\ref{3120}) it will then be possible to express  these variations in terms of the parameters $\alpha_{1}$ and $\alpha_{2}$. We have already expressed $\rho$ and $\lambda$ in terms of $g^{ij}$ (see equation(\ref{1124})). To use (\ref{3122}) directly, similar expressions involving the inverse matrix $g_{ij}$ are required.It is easy to find that $\rho$ can be expressed as
\begin{equation}
\rho = -\frac{g_{01}}{g_{11}}
\label{3124}
\end{equation}
The transformations (\ref{3122}) then lead to 
\begin{equation}
\delta \rho = - \partial_{0} \Lambda_{1} + \rho \left( \partial_{0}
\Lambda_{0} - \partial_{1} \Lambda_{1} \right)  - \left( \rho^{2}
 - \lambda^{2} \right) \partial_{1} \Lambda_{0} + 2 \rho^{2} \partial_{1}
 \Lambda_{0} - \Lambda_{k} \partial_{k} \rho
\label{3124a}
\end{equation}
$\Lambda_{i}$ in the last equation can be substituted by $\alpha_{i}$ using the mapping (\ref{3120}). We find that the resulting expression is identical with the corresponding variation, given in (\ref{GM51}) of $\rho$ under gauge transformation.

      A similar comparison can be done for $\delta \lambda$ also. The ratio
\begin{equation}
\frac{g^{11}}{g^{00}} =\left(\rho^2 - \lambda^2\right)
\label{3125}
\end{equation}
obtained from (\ref{1119}) may be taken as the starting point. We can reduce (\ref{3125}) to $ \frac{g_{00}}{g_{11}} =\left(\rho^2 - \lambda^2\right) $. Now using (\ref{3122}) and the mapping (\ref{3120}) we get the expression of $\delta \lambda$. Again we find exact matching with (\ref{GM51}). The mapping (\ref{3120}) thus establishes complete equivalence of the gauge transformations generated by the first class constraints with the diffeomorphisms of the string. 
\section{The Membrane}
 In the above we have elaborated the interpolating Lagrangean formalism
for free strings and studied the gauge symmetry from alternative approaches.
The correspondence of the gauge transformations generated by the
first class constraints and reparametrization symmetry on the world
sheet was established. The analysis, based on the
constraints, is applicable in
 general. This will be illustrated by taking the bosonic
membrane as a concrete example. Most of the results of this section will
be possible to generalize for an arbitrary p - brane.
\subsection{The interpolating lagrangean of the membrane}

 The membrane is a two dimensional object which sweeps out a three dimensional
world volume in the D dimensional space - time in which it is embedded. We will
denote the parameters parameterizing this world volume by $\tau$, $\sigma_{1}$
and $\sigma_{2}$ which will be sometimes collectively referred by the symbol
$\xi$. The natural classical action for a membrane moving in flat space - time
is given by the integrated proper volume swept out by the membrane. This
action is of the Nambu Goto form
\begin{eqnarray}
S_{\mathrm{NG}} = - \int d^{3}\xi \sqrt{-h}
\label{ss}
\end{eqnarray}
where $h$ is the determinant of the induced metric
\begin{equation}
h_{ij} = \partial_{i} X^{\mu}\partial_{j}X_{\mu}
\label{ss1}
\end{equation}
The indices $i$ and $j$ run from 1 to 3. Note that like the string we have
kept the membrane tension implicit.
The action (\ref{ss}) is again reparametrization invariant
for which $X^{\mu}$ should transform as (\ref{113}).
The primary constraints following from the Nambu--Goto action are \cite{rbk}
\begin{eqnarray}
\Omega_{a} = \Pi_{\mu}\partial_{a}X^{\mu} \approx 0 \nonumber \\
\Omega_{3} = \frac{1}{2}\left(\Pi^2 + \bar{h}\right) \approx 0
\label{I1}
\end{eqnarray}
In the above equations $\bar{h}= det(h_{ab})$, $h_{ab} = \partial_aX^{\mu}
\partial_b X_{\mu}.$ The indices $a$, $b$ run from 1 to 2 i.e.
a,b label the spatial part of the world volume of the membrane. In fact
allowing the indices $a$, $b$ to run from 1 to p the expressions written
for the membrane action and its constraint structure carry through for a
generic p - brane.

   Since the membrane action (\ref{ss}), like the string case (\ref{111}), possesses reparametrization invariance, the canonical Hamiltonian following from the action vanishes. Thus the total Hamiltonian is only a linear combination of the constraints(\ref{I1}):
\begin{equation}
{\cal{H}} = - \rho_{a}\Pi_{\mu}\partial_{a}X^{\mu} 
            - \frac{\lambda}{2}\left(\Pi^2 + \bar{h}\right) 
\label{I2}
\end{equation}
The corresponding Polyakov form is introduced as,
\begin{equation}
S_{\mathrm{P}} = -\frac{1}{2}\int d^{3}\xi{\sqrt - g} \left(g^{ij} \partial_{i} X^{\mu}\partial_{j} X_{\mu} - 1\right)
\label{s011}
\end{equation}
The metric $g_{ij}$ are now considered as independent fields. The equivalence
of (\ref{s011}) with the NG form (\ref{ss}) can be established by
substituting the solution of $g_{ij}$ in (\ref{s011}). It is instructive to
compare (\ref{s011}) with its counterpart (\ref{1112}). There is an extra
`cosmological
term' in the action (\ref{s011}). This is necessary because the Polyakov
form of the membrane action does not have Weyl invariance. Notably, the
Polyakov metric has six independent metric coefficients only three of which
can be fixed by using the reparametrization invariances. This distinguishes
the Polyakov formalism of the membrane from its string counterpart where the
metric can be completely fixed.

    We now come to the the construction of an interpolating action for the
membrane. The first step  is to consider the Lagrange multipliers as
independent fields and write an alternative first order Lagrangian for the
membrane
\begin{equation}
{\cal{L}}_I = \Pi_{\mu}\dot{X}^{\mu} - \cal{H}\label{I3}
\end{equation}
The equation of motion for $\Pi_{\mu}$ following from the Lagrangean
(\ref{I3}) is
\begin{equation}
\Pi_{\mu} = -\frac{\dot{X_{\mu}}+\rho_a\partial_{a}X^{\mu}}{\lambda}
\label{I4}
\end{equation}
Substituting $\Pi_{\mu}$ in (\ref{I3}) the form of the
Lagrangean (\ref{I3}) becomes
\begin{eqnarray}
{\cal{L}}_I& = &-\frac{\dot{X}_{\mu}+\rho_{a}\partial_{a}X_{\mu}}{\lambda}
\dot{X}^{\mu}+\frac{\lambda}{2}\left[(-\frac{\dot{X}^{\mu}
         +\rho_{a}\partial_{a}X^{\mu}}{\lambda})(-\frac{\dot{X}_{\mu}
            +\rho_{a}\partial_{a}X_{\mu}}{\lambda})+\bar{h}\right]\nonumber\\
         &  &\qquad \qquad +\rho_{a}(-\frac{\dot{X_{\mu}}+\rho_{a}\partial_{a}X_{\mu}}{\lambda})\partial_{a}X^{\mu}\label{I44}
\end{eqnarray}
Simplifying (\ref{I44}) we get the interpolating Lagrangian
\begin{equation}
{\cal{L}}_I =-\frac{1}{2\lambda}\left[\dot{X^{\mu}}\dot{X_{\mu}}
                 + 2\rho_{a}\dot{X}_{\mu}\partial_{a}X^{\mu}
                 +\rho_{a}\rho_{b}\partial_{a}X^{\mu}\partial_{b}X_{\mu}\right]
                 +\frac{\lambda}{2}\bar{h}
\label{I5}
\end{equation}
for the membrane. In the following analysis we will often check our results
for the membrane by going over to the string limit.In case of the string which
is a 1-brane, $ a,b = 1$. Then $\bar{h} = {\rm{det}}(h_{ab}) = \partial_{\sigma}X^{\mu}\partial_\sigma X_{\mu}= {X}^{\prime 2}_{\mu}$. It is easy to see that with these substitutions the Lagrangean (\ref{I5}) becomes identical with the corresponding Lagrangian (\ref{1116}) of the string. We have anticipated the name interpolating Lagrangean from our experience in the string case. Below, we will establish this by generating both the NG and the Polyakov forms of the membrane action from (\ref{I5}).
\subsection{The reduction of the interpolating lagrangean to the NG
and to the Polyakov form}
Let us first discuss the passage to the NG form. From the interpolating Lagrangean it is easy to write down the equations of motion for $\lambda$ and $\rho_a$. The Euler - Lagrange equation for $\lambda$ is
\begin{equation}
\frac{1}{2\lambda^2}\left[\dot{X_{\mu}}\dot{X^{\mu}}
       + 2\rho_a\dot{X_{\mu}}\partial_aX^{\mu}
       + \rho_a\rho_b\partial_aX_{\mu}\partial_bX^{\mu}\right]
       + \frac{1}{2}\bar{h} = 0\label{NG1}
\end{equation}
and that for $\rho_a$ are
\begin{equation}
\partial_{a}X_{\mu}\partial_{b}X^{\mu} \rho_{b}
         = -\dot{X}_{\mu}\partial_{a}X^{\mu}
\label{NG2}
\end{equation}
From the last equation we can solve $\rho_{a}$
\begin{equation}
\rho_{a} = -h_{0b}\bar h^{ba}
\label{NG3}
\end{equation}
where $h_{ab}$ has been defined below equation (\ref{I1}) and $\bar h^{ab}$ \footnote
{Note that $\bar h^{ab}$ is different from the space part of $h^{ij}.$} 
is the inverse matrix of $h_{ab}$.
Using (\ref{NG3}) in (\ref{NG1}) we get after some calculations
\begin{equation}
\lambda = -\frac{\sqrt{-h}}{\bar{h}}
\label{NG4}
\end{equation}
where we take the negative sign due to similar reason as in the string case. Substituting $\rho_a$ and $\lambda$ in (\ref{I5}) we retrieve the Nambu--Goto action. The reduction is completely analogous to the string case. In fact the solutions to $\rho_a$ and $\lambda$ go to the corresponding solutions of the string case when only one spatial degree of freedom is retained in the brane volume.

    Already in the string case the reduction of the interpolating action
to the Polyakov form was non trivial.
In case of the membrane it is further complicated by a mismatch in the
number of degrees of freedom count. The Polyakov action of the membrane
contains six independent metric components. Thus there are six more independent fields apart from $X^{\mu}$. In contrast, the interpolating Lagrangean contain only three additional fields ($\rho_1$, $\rho_2$ and $\lambda$). This mismatch is important for the choice of gauge fixing conditions in the Polyakov theory \cite{t,rbk}. Here it affects the simulation of the Polyakov Lagrangean from the interpolating Lagrangean. Understandably, this mismatch will be more pronounced for higher branes. It may be noted in this context that there is no such mismatch in the string case.

 Coming back to the problem of simulation of the Polyakov form from the
interpolating action of the membrane, it will thus be required to introduce
$(6 - 3 =)\,3$ arbitrary variables to get the Polyakov Lagrangean from the
interpolating one. The interpretation of such variables will then
be investigated self - consistently.

    We, therefore, modify (\ref{I5}) as
\begin{eqnarray}
{\cal{L}}_I & =& -\frac{1}{2\lambda}\left[\dot{X^{\mu}}\dot{X_{\mu}}
                 + 2\rho_a\dot{X_{\mu}}\partial_aX^{\mu}
              +\left(\rho_a\rho_b\partial_aX^{\mu}\partial_bX_{\mu}
                 - \lambda^{2} S_{ab}\partial_{a}X_{\mu} \partial_{b} X^{\mu}
                   \right)\right]\nonumber\\
     &&   -  \frac{\lambda}{2}\left(S_{ab}
             \partial_{a} X_{\mu} \partial_{b} X^{\mu}
     - \bar{h} - \rm{det S}\right) -\frac{\lambda}{2}\rm{det S}
\label{NG5}
\end{eqnarray}
Here $S_{ab}$ is a $2\times 2$ symmetric matrix whose elements are arbitrary
functions of $\xi^i$, the parameters labeling the membrane volume. Note
that we have introduced as many arbitrary functions which are needed to
match the extra number of degrees of freedom as mentioned above. Now
exploiting the arbitrariness of the functions $S_{ab}$ we demand that
they be chosen to satisfy
\begin{equation}
S_{ab} \partial_{a} X_{\mu} \partial_{b}X^{\mu} - \bar{h} - {\rm{det}}S = 0
\label{NG6}
\end{equation}
The condition (\ref{NG6}) can be written in a suggestive form if we substitute 
\begin{equation}
S_{ab} = \epsilon_{ac} \epsilon_{bd} G_{cd} 
\label{NG61}
\end{equation}
It is easy to check that ${\rm{det}}S = {\rm{det}}G $. Using (\ref{NG61}), the condition (\ref{NG6}) can be cast as
\begin{equation}
\mathrm{det}\left(G_{ab} - h_{ab}\right) = 0
\label{NG66}
\end{equation}
We observe that this condition is reminiscent of a weaker version of the first class constraint $g_{ab} = h_{ab}$ of the Polyakov action. In the following we will find that this coincidence is not accidental. Now substituting $S_{ab}$ by $G_{ab}$ in (\ref{NG5}) we get
\begin{eqnarray}
{\cal{L}}_I & =& -\frac{1}{2\lambda}\left[\dot{X^{\mu}}\dot{X_{\mu}}
                 + 2\rho_a\dot{X_{\mu}}\partial_aX^{\mu}
              +\left(\rho_a\rho_b\partial_aX^{\mu}\partial_bX_{\mu}
                 - \lambda^{2} \epsilon_{ac} \epsilon_{bd} G_{cd} \partial_{a}
                 X_{\mu} \partial_{b} X^{\mu}\right)\right]\nonumber\\
     &&   -  \frac{\lambda}{2}\left(\epsilon_{ac} \epsilon_{bd} G_{cd}
     \partial_{a} X_{\mu} \partial_{b} X^{\mu}
     - \bar{h} - \rm{det G}\right) -\frac{\lambda}{2}\rm{det G}
\label{NG51}
\end{eqnarray}
It is now possible to reduce equation (\ref{NG51}) in the form
\begin{equation}
{\cal{L}}_I = -\frac{1}{2}\sqrt{-g} g^{ij}\partial_iX_{\mu}
               \partial_jX^{\mu} - \frac{\lambda}{2} {\rm{det}} G
\label{NG7}
\end{equation}
where
\begin{equation}
g^{ij} = \left(-g\right)^{-\frac{1}{2}}
\left(\begin{array}{cc}
\frac{1}{\lambda} & \frac{\rho_a}{\lambda} \\
\frac{\rho_a}{\lambda} & \frac{\rho_{a} \rho_{b} - \lambda^2 \epsilon_{ac}\epsilon_{bd}
G_{cd} }{\lambda} \\
\end{array}\right)
\label{idm}
\end{equation}
 From the above identification we get after a straightforward calculation that
\begin{equation}
{\rm{det}} g^{ij} = \frac{\lambda {\rm{det}}G }{\left(-g\right)^{\frac{3}{2}}}
\label{NG8}
\end{equation}
But we require ${\rm {det}} g^{ij} $ = $ g^{-1} $. Comparing, we get the
condition
\begin{equation}
\lambda {\rm{det}} G = -\sqrt{-g}
\label{NG81}
\end{equation}
Using the above condition in (\ref{NG7}) we find
\begin{equation}
{\cal{L}}_I = -\frac{1}{2}\sqrt{-g}\left(g^{ij}\partial_{i} X_{\mu}
\partial_{j} X^{\mu} - 1 \right)
\label{NG9}
\end{equation}
which is the Polyakov version of the membrane action. Note that the
cosmological term appears automatically in this simulation. This is a new
feature for the membrane which was not present in the analogous construction
for the string.

    At this point it is appropriate to check the consistency of the above
construction as we did in the string case. Referring to the identification (\ref{idm}) we find from the expressions of $g_{00}$ and $g_{0a}$
\begin{eqnarray}
\frac{1}{\lambda} &=& \sqrt{-g} g^{00}\nonumber\\
\rho_{a}&=& \frac{g^{0a}}{g^{00}}
\label{c1}
\end{eqnarray}
From the above expressions we can solve $\rho_{a}$ and $\lambda$ in terms of the appropriate elements of $g^{ij}$. Since $\rho_{a}$ and $\lambda$ occur in different specific combinations in the space part 
\begin{equation}
g^{ab} = \left(-g\right)^{-\frac{1}{2}}
         \frac{\rho_{a} \rho_{b} - \lambda^2 \epsilon_{ac}\epsilon_{bd}G_{cd} }         {\lambda} 
\label{idm1}
\end{equation}
 it is necessary to see what happens when the above solutions of $\rho_{a}$ and $\lambda$ are substituted in (\ref{idm1}). In particular, from
\begin{equation}
g ^{11} = \frac{1}{\sqrt {- g}}\frac{\rho_1^2 - \lambda^{2} G_{22}}{\lambda}
\label{c2}
\end{equation}
we get after some manipulations 
\begin{equation}
G_{22} = \frac{g^{00}g^{11} - (g^{01})^2}{g^{-1}} = g_{22}\label{c3}
\end{equation}
Similarly, starting from the remaining terms of (\ref{idm1}) we arrive at
\begin{equation}
G_{ab} = g_{ab}
\label{c4}
\end{equation}
The arbitrary functions $G_{ab}$ introduced earlier are thus identified with
the spatial part of $g_{ij}$. Note that this coincidence is due to the special choice of the arbitrary functions (\ref{NG61}). From (\ref{c4}) we observe that equation (\ref{NG66}) is really the weaker version of the first class constraint $g_{ab} = h_{ab}$ following from the Polyakov action. Further, from (\ref{c4}) we get 
\begin{equation}
{\rm{det}}G = \bar g
\label{c5}
\end{equation}
where $\bar g$ is the determinant of $g_{ab}$. The solution of $\lambda$ from (\ref{c1}) is 
\begin{eqnarray}
\lambda = \frac{1}{\sqrt{-g} g^{00}}
\label{c6}
\end{eqnarray}
Hence we can calculate $\lambda {\rm{det}}G$ as 
\begin{eqnarray}
\lambda {\rm{det}}G = \frac{1}{\sqrt{-g} g^{00}} \bar g
\label{c7}
\end{eqnarray}
But $\bar g = g g^{00}$. Substituting this in (\ref{c7}) we see that the value of $\lambda {\rm{det}}G$ is the same as (\ref{NG81}). Therefore, the identification (\ref{c4}) is consistent with (\ref{NG81}). Finally, one may enquire whether the form of $G_{ab}$ given by (\ref{c4}) is consistent with direct computation of the inverse of (\ref{idm}). It is indeed gratifying to observe that the space part of the inverse matrix coincides with $G_{ab}$. The consistency of the construction (\ref{idm}) is thus completely verified.

 Finally, it will be instructive to explore the connection between (\ref{idm}) and the metric in ADM representation for $d = 3$. In fact, we shall establish the exact mapping between them. The ADM metric for $d = 3$ assumes the form
\begin{equation}
\left(\begin{array}{ccc}
{}^{\left(3\right)}\gamma^{00}&{}^{\left(3\right)}\gamma^{01}&{}^{\left(3\right)}\gamma^{02}\\
{}^{\left(3\right)}\gamma^{01}&{}^{\left(3\right)}\gamma^{11}&{}^{\left(3\right)}\gamma^{12}\\
{}^{\left(3\right)}\gamma^{02}&{}^{\left(3\right)}\gamma^{12}&{}^{\left(3\right)}\gamma^{22}\\
\end{array}\right)
 =\left(\begin{array}{ccc}
-\frac{1}{\left(N\right)^{2}}&\frac{\left(N^{1}\right)}{\left(N\right)^{2}}&\frac{\left(N^{2}\right)}{\left(N\right)^{2}}\\
\frac{\left(N^{1}\right)}{\left(N\right)^{2}}&\left(\gamma^{11}-\frac{\left(N^{1}\right)^{2}}{\left(N\right)^{2}}\right)&\left(\gamma^{12}-\frac{\left(N^{1}\right)\left(N^{2}\right)}{\left(N\right)^{2}}\right)\\
\frac{\left(N^{2}\right)}{\left(N\right)^{2}}&\left(\gamma^{12}-\frac{\left(N^{1}\right)\left(N^{2}\right)}{\left(N\right)^{2}}\right)&\left(\gamma^{22}-\frac{\left(N^{2}\right)^{2}}{\left(N\right)^{2}}\right)\\
\end{array}\right)
\label{ADM8}
\end{equation}
This involves one lapse variable $N$, two shift variables $N^{k}$, ($k = 1, 2$) and the metric $\gamma^{ab}$, ($a,b = 1, 2$) on the two dimensional hypersurface. We will compare (\ref{ADM8}) with (\ref{idm}). If we introduce the connection \begin{eqnarray}
\left(N^{1}\right)\mapsto -\rho_{1},\quad\left(N^{2}\right)\mapsto -\rho_{2} \quad \mathrm {and}\quad \left(N\right)^{2}\mapsto -\lambda\sqrt{-g}
\label{ADM5}
\end{eqnarray}
and impose the condition 
\begin{equation}
\gamma^{ab} = -\lambda \left(-g\right)^{-\frac{1}{2}}
\left(\begin{array}{cc}
G_{22} & -G_{12} \\
-G_{12} & G_{22} \\
\end{array}\right)
\label{ADM6}
\end{equation}
then the $d = 3$ ADM metric (\ref{ADM8}) goes over to our construction (\ref{idm}). The inverse of $\gamma^{ab}$ can be calculated from (\ref{ADM6}) as
\begin{equation}
\gamma_{ab} = - \frac{\left(-g\right)^{\frac{1}{2}}}{\lambda \mathrm{det} G}
G_{ab}
\label{ADM7}
\end{equation}
Using (\ref{NG81}) we get 
\begin{equation}
\gamma_{ab} = G_{ab}
\label{ADM9}
\end{equation}
 This connection with the ADM representation also shows that the metric on the hypersurface $\gamma_{ab}$ is just the arbitrary $2\times 2$ symmetric matrix $G_{ab}$ which we introduced earlier in this section to match the extra degrees of freedom. Consequently, the totally arbitrary nature of $\gamma_{ab}$ is confirmed. In this connection, note that the corresponding metric $\gamma^{11}$ in the case of the string was fixed by exploiting the Weyl invariance.

\subsection{ Gauge Symmetry of the Membrane}
The investigation of the gauge symmetry of the interpolating 
membrane can be pursued following essentially the same steps as in the
string case discussed earlier in section-$3$. There we argued that the gauge
variations of the fields $\rho$ and $\lambda$ can be obtained using the
symmetries of the N-G action, where they appear as Lagrange multipliers, by
applying the formula (\ref{218}). The same arguments also apply here. So we
 construct the generator of the gauge transformations as
\begin{equation}
G = \int d\xi \alpha_{i}\left( \xi \right) \Omega_{i}
\left( \xi \right) ; \left(i= 1,2,3 \right)
\label{GM1}
\end{equation}
where $\Omega_{i}$ are the constraints given in 
(\ref{I1}), and $ \alpha_{i}\left( \xi \right) $ are the three arbitrary
gauge parameters.
The algebra of the constraints can be worked out using (\ref{I1})
\begin{eqnarray}
\left\{ \Omega_{a}\left( \xi \right), \Omega_{b}\left( \xi^{\prime} \right)
\right\} &=& \left[\Omega_{b}\left( \xi \right) \partial_{a}\left(\delta
\left(\xi - \xi^{\prime} \right)\right)
- \Omega_{a}\left( \xi^{\prime} \right)  \partial_{b}^{\prime} \left(\delta
\left(\xi - \xi^{\prime} \right)\right)\right] \nonumber \\
\left\{ \Omega_{a}\left( \xi \right), \Omega_{3}\left( \xi^{\prime} \right)
\right\} &=& \left[\Omega_{3}\left( \xi \right) + \Omega_{3}\left( \xi^{\prime}
\right)\right] \partial_{a}\left(\delta\left(\xi - \xi^{\prime}\right)\right)
\nonumber \\
\left\{ \Omega_{3}\left( \xi \right), \Omega_{3}\left( \xi^{\prime} \right)
\right\} &=& 4 \left[ \bar {h}\left( \xi \right)\bar {h}^{ab}
\left( \xi \right)\Omega_{b}\left( \xi \right) \right.\nonumber \\
&& + \left.\bar {h}\left( \xi^{\prime} \right)\bar {h}^{ab}\left( \xi^{\prime} \right)
\Omega_{b}\left( \xi^{\prime} \right) \right]
\partial_{a}\left(\delta\left(\xi - \xi^{\prime}\right)\right)
\label{d1}
\end{eqnarray}
From this algebra we read off the non-zero structure functions as defined in (\ref{315}),
\begin{eqnarray}
C_{11}{}^{1}\left( \xi, \xi^{\prime}, \xi^{\prime \prime} \right)
&=& \left[ \delta\left( \xi - \xi^{\prime \prime} \right) +
\delta\left( \xi^{\prime} - \xi^{\prime \prime} \right) \right]
\partial_{1}\left(\delta \left( \xi - \xi^{\prime} \right) \right) \nonumber \\
C_{12}{}^{1}\left( \xi, \xi^{\prime}, \xi^{\prime \prime} \right)
&=& \delta\left( \xi^{\prime} - \xi^{\prime \prime} \right)
\partial_{2}\left(\delta \left( \xi - \xi^{\prime} \right) \right) \nonumber \\
C_{12}{}^{2}\left( \xi, \xi^{\prime}, \xi^{\prime \prime} \right)
&=& \delta\left( \xi - \xi^{\prime \prime} \right)
\partial_{1}\left(\delta \left( \xi - \xi^{\prime} \right) \right) \nonumber \\
C_{21}{}^{1}\left( \xi, \xi^{\prime}, \xi^{\prime \prime} \right)
&=& \delta\left( \xi - \xi^{\prime \prime} \right)
\partial_{2}\left(\delta \left( \xi - \xi^{\prime} \right) \right) \nonumber \\
C_{21}{}^{2}\left( \xi, \xi^{\prime}, \xi^{\prime \prime} \right)
&=& \delta\left( \xi^{\prime} - \xi^{\prime \prime} \right)
\partial_{1}\left(\delta \left( \xi - \xi^{\prime} \right) \right) \nonumber \\
C_{22}{}^{2}\left( \xi, \xi^{\prime}, \xi^{\prime \prime} \right)
&=& \left[ \delta\left( \xi - \xi^{\prime \prime} \right) +
\delta\left( \xi^{\prime} - \xi^{\prime \prime} \right) \right]
\partial_{2}\left(\delta \left( \xi - \xi^{\prime} \right) \right) \nonumber \\
C_{13}{}^{3}\left( \xi, \xi^{\prime}, \xi^{\prime \prime} \right)
&=& C_{31}{}^{3}\left( \xi, \xi^{\prime}, \xi^{\prime \prime} \right)
= \left[ \delta\left( \xi - \xi^{\prime \prime} \right) +
\delta\left( \xi^{\prime} - \xi^{\prime \prime} \right) \right]
\partial_{1}\left(\delta \left( \xi - \xi^{\prime} \right) \right) \nonumber \\
C_{23}{}^{3}\left( \xi, \xi^{\prime}, \xi^{\prime \prime} \right)
&=& C_{32}{}^{3}\left( \xi, \xi^{\prime}, \xi^{\prime \prime} \right)
= \left[ \delta\left( \xi - \xi^{\prime \prime} \right) +
\delta\left( \xi^{\prime} - \xi^{\prime \prime} \right) \right]
\partial_{2}\left(\delta \left( \xi - \xi^{\prime} \right) \right) \nonumber \\
C_{33}{}^{1}\left( \xi, \xi^{\prime}, \xi^{\prime \prime} \right)
&=& 4 \left[ \bar {h}\left( \xi \right) \bar {h}^{11}\left( \xi \right)
\partial_{1}\left(\delta\left(\xi - \xi^{\prime}\right)\right) \right.\nonumber \\
&& \qquad + \left. \bar {h}\left( \xi \right) \bar {h}^{21}\left( \xi \right)
\partial_{2}\left(\delta\left(\xi - \xi^{\prime}\right)\right) \right]
 \delta \left(\xi -\xi^{\prime \prime} \right)\nonumber \\
&&+ 4 \left[ \bar {h}\left( \xi^{\prime} \right)\bar {h}^{11}
\left( \xi^{\prime} \right) \partial_{1}\left(\delta\left(\xi - \xi^{\prime}
\right)\right) \right. \nonumber \\
&& \qquad + \left. \bar {h}\left( \xi^{\prime} \right)\bar {h}^{21}
\left( \xi^{\prime} \right) \partial_{2}\left(\delta\left(\xi - \xi^{\prime}
\right)\right) \right]\delta \left(\xi^{\prime}  - \xi^{\prime \prime}
\right) \nonumber \\
C_{33}{}^{2}\left( \xi, \xi^{\prime}, \xi^{\prime \prime} \right)
&=& 4 \left[ \bar {h}\left( \xi \right) \bar {h}^{12}\left( \xi \right)
\partial_{1}\left(\delta\left(\xi - \xi^{\prime}\right)\right) \right.\nonumber \\
&& \qquad + \left. \bar {h}\left( \xi \right) \bar {h}^{22}\left( \xi \right)
\partial_{2}\left(\delta\left(\xi - \xi^{\prime}\right)\right) \right]
 \delta \left(\xi -\xi^{\prime \prime} \right)\nonumber \\
&&+ 4 \left[ \bar {h}\left( \xi^{\prime} \right)\bar {h}^{12}
\left( \xi^{\prime} \right) \partial_{1}\left(\delta\left(\xi - \xi^{\prime}
\right)\right) \right. \nonumber \\
&& \qquad + \left. \bar {h}\left( \xi^{\prime} \right)\bar {h}^{22}
\left( \xi^{\prime} \right) \partial_{2}\left(\delta\left(\xi - \xi^{\prime}
\right)\right) \right]\delta \left(\xi^{\prime}  - \xi^{\prime \prime}
\right) \nonumber \\
\label{d2}
\end{eqnarray}
Relabeling $\rho_{a}$ and $\lambda$ as
\begin{equation}
\lambda_{a} = \rho_{a} \hspace{1cm} \rm{and} \hspace{1cm}
\lambda_{3} = \frac{\lambda}{2}
\label{d3}
\end{equation}
and using the structure functions (\ref{d2}) we calculate the required
gauge variations by applying equation (\ref{314})
\begin{eqnarray}
\delta \lambda_{a} &=& - \dot \alpha_{a}
+ \left(\alpha_{b}\partial_{b}\lambda_{a}
 - \lambda_{b}\partial_{b}\alpha_{a} \right) + 4 \bar{h}\bar{h}^{ab}
\left(\alpha_{3}\partial_{b}\lambda_{3} - \lambda_{3}\partial_{b}
\alpha_{3}\right)\nonumber\\
\delta \lambda_{3} &=& -\dot \alpha_{3}
+\left(\alpha_{3}\partial_{a}\lambda_{a} - \lambda_{a} \partial_{a}\alpha_{3}
\right)
+\left(\alpha_{a}\partial_{a}\lambda_{3} - \lambda_{3} \partial_{a}\alpha_{a}
\right)
\label{GM4}
\end{eqnarray}
Now we use equation (\ref{d3}) to convert $\lambda_{i}$ back to
 $\rho_1$, $\rho_2$ and $\lambda$
\begin{eqnarray}
\delta \rho_{a} &=& - \dot \alpha_{a}
+ \left(\alpha_{b}\partial_{b}\rho_{a}
 - \rho_{b}\partial_{b}\alpha_{a} \right) + 2 \bar{h}\bar{h}^{ab}
\left(\alpha_{3}\partial_{b}\lambda - \lambda\partial_{b}\alpha_{3}\right)
\nonumber\\
\delta \lambda &=& -2 \dot \alpha_{3}
+2 \left(\alpha_{3}\partial_{a}\rho_{a} - \rho_{a} \partial_{a}\alpha_{3}
\right)
+\left(\alpha_{a}\partial_{a}\lambda - \lambda \partial_{a}\alpha_{a}\right)
\label{GM5}
\end{eqnarray}
Note that the above variations of $\rho_{a}$ and $\lambda$ can be directly obtained from the definitions (\ref{NG3}) and (\ref{NG4}) in complete parallel with the analogous computation for string (see under equation (\ref{GM51})). Also, it is instructive to study the string limit of equations (\ref{GM5}). In the string limit we put $a = b = 1$. The matrix $h_{ab} = \partial_{a}X^{\mu}\partial_{b}X_{\mu}$ now contains only one term, namely $ h_{11}$. So $\bar h^{ab}$ now also contains a single term $ h^{11} = \frac{1}{h_{11}}$. Hence in the string limit $\bar h \bar h^{ab}$ becomes $h_{11} \times \frac{1}{h_{11}}$ i.e. $1$.  It is now apparent that in the string limit we recover the expressions (\ref{GM51}) from (\ref{GM5}) with the replacement of $\alpha_{3}$ by $\alpha_{2}$.
   Before concluding this section we will investigate the parallel between
gauge symmetry and reparametrization symmetry of the membrane actions using
our alternative approaches developed in the string example.
We, therefore, require first to find a correspondence between the
transformation parameters in both the cases. This may be obtained by comparing
the transformations on $X^{\mu}$. The variations of $X^{\mu}$
in (\ref{I5}) under (\ref{GM1}) is
\begin{equation}
\delta X_{\mu} = \left\{X_{\mu}, G \right\}
 = \left( \alpha_{a} \partial_{a} X_{\mu} + 2 \alpha_{3} \Pi_{\mu} \right)
\label{GM7}
\end{equation}
 Looking at the scenario from the point of view of Polyakov action we find
that under reparametrization the variations of $X^{\mu}$ and $g_{ij}$ follow
from relations similar to (\ref{3117}) and (\ref{3122}). Thus
\begin{equation}
\delta X^{\mu} = \Lambda^{i} \partial_{i} X^{\mu}
= \Lambda^{0} \dot X^{\mu} + \Lambda^{a} \partial_{a} X^{\mu}
\label{GM8}
\end{equation}
Using (\ref{I4}) we eliminate $\dot X^{\mu}$ in terms of the momenta $\Pi^{\mu}$. Now comparing (\ref{GM8}) with (\ref{GM7}) we obtain the mapping
\begin{eqnarray}
\Lambda^{0} &=& - 2 \sqrt{- g } g^{00} \alpha_{3}    = - \frac{ 2 \alpha_{3}}{\lambda}
\nonumber \\
\Lambda^{a} &=& \alpha_{a} - 2 \sqrt{- g } g^{0a} \alpha_{3}    = \alpha_{a} -
\frac{ 2 \rho_{a} \alpha_{3}}{\lambda}
\label{GM9}
\end{eqnarray}
Under (\ref{GM9}) the transformations on $X^{\mu}$ due to reparametrization become identical with its corresponding gauge variation.
The complete equivalence between the transformations can again be established by computing $\delta \rho_{a}$ and $\delta \lambda$ from the alternative approach.The mapping (\ref{idm}) yields,
\begin{equation}
\rho_{a} = \frac{g^{0a}}{g^{00}}
\label{GM11}
\end{equation}
We require to express these in terms of $g_{ij}$. To this end we start
from the identity
\begin{equation}
g^{ij} g_{jk} = \delta^{i}{}_{k}
\label{GM12}
\end{equation}
and obtain the following equations for $\rho_{a}$
\begin{eqnarray}
\rho_{1} g_{11} + \rho_{2} g_{21} = - g_{01}\nonumber \\
\rho_{1} g_{12} + \rho_{2} g_{22} = - g_{02}
\label{GM13}
\end{eqnarray}
Solving the above equations we can express $\rho_{a}$ entirely in terms of
$g_{ij}$. The variations of $g_{ij}$ under reparametrization is obtained from
(\ref{3122}) where $i,j$ now assume values $0$, $1$ and $2$. So the
corresponding variations of $\rho_{a}$ are given by
\begin{eqnarray}
\delta \rho_{a} &=& - \partial_{0} \Lambda^{a} +
\rho_{a} \partial_{0} \Lambda^{0} - \rho_{b} \partial_{b} \Lambda^{a} 
 + \rho_{a}\rho_{b} \partial_{b} \Lambda^{0}
+ \Lambda^{k}\partial_{k} \rho_{a}\nonumber \\
&& \qquad\qquad\qquad - \lambda^{2}\epsilon_{ab}\epsilon_{cd} g_{bc}
\partial_{d} \Lambda^{0}
\label{GM14}
\end{eqnarray}
Now introducing the mapping (\ref{GM9}) in (\ref{GM14}) and using the first class constraints $g_{ab} = h_{ab}$ we find that the variations of $\rho_{a}$ are identical with their gauge variations in (\ref{GM5}). We then compute the variation of $\delta\lambda$. This can be conveniently done by starting from the variation of the ratio
\begin{equation}
\frac{g^{11}}{g^{00}} =\left(\rho_{1}^{2} - \lambda^2 g_{22}\right)
\label{3125A}
\end{equation}
obtained from the identification (\ref{idm}). Converting the l.h.s appropriately in terms of $g_{ij}$, we take the gauge variation to get
\begin{equation}
\delta \left\{\frac{g^{11}}{g^{00}}\right\} = \delta\left\{\frac{g_{00}g_{22} - g_{02}{}^{2}}{g_{11}g_{22} - g_{12}{}^{2}}\right\}= \delta \left(\rho_{1}^{2} - \lambda^2 g_{22}\right)
\label{3125B}
\end{equation}
and using the variations (\ref{3122}) and (\ref{GM14}) we get the expression of $\delta \lambda$ in terms of the reparametrization parametres $\Lambda_{i}$. Using the mapping (\ref{GM9}) we substitute $\Lambda_{i}$ by $\alpha_{i}$ and  the resulting expression for $\delta \lambda$ can be easily shown to agree with that given in (\ref{GM5}). The complete matching thus obtained illustrates the equivalence of reparametrization symmetry with gauge symmetry for the membrane.

\section{Conclusion}
  We have developed a new action formalism for bosonic membranes and
demonstrated that it interpolates between the Nambu--Goto and the
Polyakov forms of the membrane actions. This is similar to the interpolating
action formalism for strings recently proposed in \cite{rbs}. The analysis
for the membrane, however, revealed some interesting aspects which
differ from string. These differences originate from the mismatch of the number of independent metric components with the number of independent reparametrizations in the membrane problem. A definite  number of arbitrary variables that properly accounted for the mismatch were required to be introduced in
the interpolating Lagrangean to reduce it to the Polyakov form. The internal
consistency of the construction was demonstrated. The precise mechanism of the introduction of independent metric on the string and membrane world volume was revealed by the interpolating Lagrangean formalism. Its connection with the ADM representation of general relativity was discussed. The lapse and shift variables in this representation were identified with the Lagrange multipliers enforcing the pair of first class constraints in strings or higher brains. Moreover, the arbitrary metric ($\gamma_{ab}$)  in the ADM picture exactly corresponds to the arbitrary metric ($G_{ab}$) introduced for interpolating between the Nambu--Goto and Polyakov type actions. A remarkable feature of our analysis was the emergence of the cosmological term in the Polyakov action from the consistency conditions. The significance of the interpolating action formalism was thereby revealed. A thorough analysis of the gauge symmetries of interpolating actions for strings and membranes was performed using a general method \cite{brr} based on Dirac's theory of constrained Hamiltonian analysis \cite{dir}. Specifically we have demonstrated the equivalence of the reparametrization invariances of different string and membrane actions with the gauge invariances generated by the first class constraints. Indeed, the whole analysis of the interpolating Lagrangean formalism was based on the local gauge symmetries only. The appearance (or otherwise) of the Weyl invariance was a logical consequence of this construction. It therefore fits into the general scheme of the recent work \cite{thi} where the loop quantum gravity quantization of string was based only on the local diffeomorphism invariance.                                              
\section*{Acknowledgment}
One of the authors (AS) wants to thank the Council of Scientific and Industrial Research (CSIR), Govt. of India, for financial support and the Director, S. N. Bose National Centre for Basic Sciences, for providing computer facilities. 

\end{document}